# MAGNETO-OPTICS OF MULTI-WELL QUANTUM STRUCTURES WITH $D_2^-$-CENTRES

V.D. Krevchik[1, 2], A.B. Grunin[1], V.V. Evstifeev[1],
M.B. Semenov[1, 2], A.K. Aringazin[2, 3]

[1] Physics Department, Penza State University, Penza, Russia,
physics@diamond.stup.ac.ru
[2] Institute for Basic Research, P.O. Box 1577, Palm Harbor,
FL 34682, USA
ibr@verizon.net
[3] Institute for Basic Research, Eurasian National University,
Astana 010008, Kazakhstan
aringazin@mail.kz

The problem of electron binding states in field of two $D^0$ – centres in semiconductive quantum well (QW) in the presence of an external longitudinal magnetic field (along the QW growth axis) is studied within the framework of zero-range potential model. It is found that the magnetic field leads to a considerable change in positions of $g$ - and $u$ -terms, and to a stabilization of the $D_2^-$-states in QW. It is shown that a form of impurity magneto-optical absorption spectrum essentially depends on the light polarization direction and on the spatial configuration of the $D_2^-$ molecular ion in QW.

## 1. Introduction

As it has been experimentally shown [1], reactions of the type $D^0 + e \to D^-$ are possible in low-dimensional systems under certain conditions. As the result of such reactions, neutral small donors bind an additional electron with the population formation of the so called "$D^-$-states". Such states, which are confined by the structure potential, open new possibilities to study correlation effects in low-dimensional systems [1].

In the present paper we consider the following specific case: $D^0$-positions can not all be effectively filled by an electron transfer through a barrier [2]. In this case, a formation of the negative molecular ion $D_2^-$ is possible; this depends on the distance $R$ between the $D^0$-centers. It should be noted that the system consisting of a weakly bounded electron in the field of two equivalent potential centers appears in alkaline-halloids crystals [3]. This is the so called "dyeing center" $M^-$, which is electron in the field of neutral $M$-center (two neighbor $F$-centers). As it is known [4, 5], $D^-$-center is a simplest system, which can be simulated by an electron in the field of the zero-range potential. Earlier we have shown [6, 7] that the zero-range potential method allows to obtain analytical solution to the wave function and the binding energy of electron, which is localized on $D^0$-center. This method allows also to investigate the impurity magneto-optical absorption in nanostructures with a parabolic confinement potential.

The negative molecular ion $D_2^-$ simulation and investiogfation of its magneto-optical properties in QW are of much interest. Since $D_2^-$-system is symmetric with respect to the center, electron states (at fixed distance $R$ between $D^0$-centers) should be either symmetric ($g$-terms) or anti-symmetric ($u$-terms). Evidently, splitting of the $g$- and $u$-terms (which are degenerated at large $R$) is determined by the value of $R$ and by the QW-parameters (as a consequence of the dimensional reduction). On the other hand, magnetic field (which is applied along the QW growth axis) plays the role of a variable parameter, which can change the system's geometric confinement and, hence, to control energies of optical transitions. This gives some perspectives to develop molecular electronics, particularly, one-molecule devices with controllable characteristics.

The aim of this work is to investigate magneto-optical spectrum of QW within the framework of the zero-potential model. Structure of the spectrum is related to electron optical transitions from $g$-term state to QW hybrid-quantized states, depending on the light polarization direction and on the spatial configuration of the $D_2^-$-molecular ion.

## 2. $D_2^-$-molecular ion terms

To describe QW one-electron states one can use the parabolic confinement potential of the form

$$U(z) = \frac{m^* \omega_0^2 z^2}{2}, \qquad (1)$$

where $m^*$ is the electron effective mass, $|z| \leq L/2$, $L$ is the width of QW, $\omega_0$ is the confinement potential characteristic frequency of QW.

For one-electron states (which are undisturbed by impurities) under the longitudinal magnetic field $\vec{B}(0,0,B)$, Hamiltonian of the model is of the following form:

$$\hat{H} = -\frac{\pi^2}{2m^*}\left(\frac{1}{\rho}\frac{\partial}{\partial \rho}\left(\rho \frac{\partial}{\partial \rho}\right) + \frac{1}{\rho^2}\frac{\partial^2}{\partial \varphi^2}\right) + \frac{\omega_B}{2}\hat{M}_z + \frac{m^* \omega_B^2 \rho^2}{8} + \hat{H}_z, \quad (2)$$

where $\omega_B = |e|B/m^*$ is the cyclotron frequency, $\hat{M}_z = -it\partial/\partial \varphi$ is the operator of orbital angular momentum projection on the $z$-axis;

$\hat{H}_z = -\left(\hbar^2/2m^*\right)\partial z/\partial z^2 + m^*\omega_0^2 z^2/2$. Double-center potential $V_\delta\left(\rho,\varphi,z;\rho_{a_1},\varphi_{a_1},z_{a_1};\rho_{a_2},\varphi_{a_2},z_{a_2}\right)$ is modeled by the superposition of the zero-range potentials with intensity $\gamma_i = 2\pi\hbar^2/(\alpha_i m^*)$ $(i = 1,2)$:

$$V_\delta\left(\rho,\varphi,z;\rho_{a_1},\varphi_{a_1},z_{a_1};\rho_{a_2},\varphi_{a_2},z_{a_2}\right) = \sum_{i=1}^{2}\gamma_i \frac{\delta(\rho-\rho_{a_i})}{\rho}\delta(\varphi-\varphi_{a_i})\delta(z-z_{a_i}) \times$$

$$\times\left[1+(\rho-\rho_{a_i})\frac{\partial}{\partial\rho}+(z-z_{a_i})\frac{\partial}{\partial z}\right]\partial z, \quad (3)$$

where $\alpha_i$ is determined by the energy $E_i = -\pi^2\alpha_i^2/(2m^*)$ of electron localized states on the same $D^-$-centers in a bulk semiconductor.

The wave function $\Psi_\lambda\left(\vec{r};\vec{R}_{a_1},\vec{R}_{a_2}\right)$ of the electron ($\vec{R}_{a_i} = \rho_{a_i},\varphi_{a_i},z_{a_i}$ is the impurity center coordinates), which is localized on $D_2^-$-centre, satisfying Lippman-Schwinger equation for the bound state, is a linear combination

$$\Psi_\lambda\left(\vec{r};\vec{R}_{a_1},\vec{R}_{a_2}\right) = \sum_{i=1}^{2}\gamma_i c_i G\left(\vec{r},\vec{R}_{a_i};E_\lambda\right),$$

(4)

where $G\left(\vec{r},\vec{R}_{a_i};E_\lambda\right)$ is one-electron Green function corresponding to the source in point $\vec{R}_{a_i}$ and to the energy $E_\lambda = -\hbar^2\lambda^2/(2m^*)$ ($E_\lambda$ is the electron binding energy in the field of $D^0$-centers under longitudinal magnetic field (this energy is measured from the bottom of QW). From mathematical point of view, the double-center problem leads to finding of nontrivial solutions of the algebraic equations for $c_i$ coefficients; this implies a transcendental equation for $E_\lambda$. In the case when $\gamma_1 = \gamma_2 = \gamma$, the latter equation is split in two equations, which determine the symmetric ($g$-term) and anti-symmetric ($u$-term) electron states. Accounting for the one-electron Green function for the case when the $D_2^-$-center axis is along the QW – growth axis $\vec{R}_{a_1}(0,0,0)$ and $\vec{R}_{a_2}(0,0,z_{a_2})$, these equations become

$$-\frac{1}{2\sqrt{\pi\beta}\eta_i}\left\{\int_0^\infty dt e^{-\left(\beta\eta_B^2+\beta a_B^{*-2}+\frac{1}{2}\right)t}\left[2^{\frac{1}{2}}\beta a_B^{*-2}\left(1-e^{-2t}\right)^{-\frac{1}{2}}\delta^{-1}(t)sh^{-1}\left(\beta a_B^{*-2}t\right)\times\right.\right.$$
$$\left.\times\left(1\pm\exp\left(-\frac{z_{a_2}^{*2}ctht}{4\beta}\right)\right)-t^{-\frac{3}{2}}\left(1\pm e^{-\frac{z_{a_2}^{*2}}{4\beta t}}\right)\right]-2\sqrt{\pi}\left(\sqrt{\beta\eta_B^2+\beta a_B^{*-2}+\frac{1}{2}}\mp\right.$$
$$\left.\left.\mp\exp\left(-\sqrt{\beta\eta_B^2+\beta a_B^{*-2}+\frac{1}{2}}\cdot\left|z_{a_2}^*\right|/\sqrt{\beta}\right)\cdot\sqrt{\beta}/\left|z_{a_2}^*\right|\right)\right\}=1,$$

(5)

where $\beta = L^*/\left(4\sqrt{U_0^*}\right)$, $L^* = L/a_\alpha$, $a_\alpha$ is the effective Bohr radius, $U_0^* = U_0/E_d$, $U_0$ is the QW confinement potential amplitude; $E_d$ is the effective Bohr energy, $\eta_B^2 = |E_\lambda|/E_d$, $\delta(t) = \exp\left(-\beta a_B^{*-2}t\right)$, $a_B^2 = a_B/a_d$, $z_{a_2}^* = z_{a_2}/a_d$; upper sign in Eq. (5) corresponds to symmetric ($g$-term), and lower sign to anti-symmetrical ($u$-term) electron states.

In the case when $D_2^-$-center axis ($\vec{R}_{a_1}(0,0,0)$ and $\vec{R}_{a_2}(\rho_{a_2},\varphi_{a_2},0)$) is transverse to the magnetic field direction, the corresponding equations can be written as

$$-\frac{1}{2\sqrt{\pi\beta}\eta_i}\left\{\int_0^\infty dt e^{-\left(\beta\eta_B^2+\beta a_B^{*-2}+\frac{1}{2}\right)t}\left[2^{\frac{1}{2}}\beta a_B^{*-2}\left(1-e^{-2t}\right)^{-\frac{1}{2}}\delta^{-1}(t)sh^{-1}\left(\beta a_B^{*-2}t\right)\times\right.\right.$$
$$\left.\times\left(1\pm\exp\left(-\frac{\rho_{a_2}^{*2}cth\left(\beta a_B^{*-2}t\right)}{4a_B^{*-2}}\right)\right)-t^{-\frac{3}{2}}\left(1\pm e^{-\frac{\rho_{a_2}^{*2}}{4a_B^{*-2}t}}\right)\right]\pm 2\sqrt{\pi}\times$$
$$\left.\times\left(\exp\left(-\sqrt{\beta\eta_B^2+\beta a_B^{*-2}+\frac{1}{2}}\cdot\frac{\rho_{a_2}^*}{a_B^*}\right)\frac{\rho_{a_2}^*}{a_B^*}\mp\sqrt{\beta\eta_B^2+\beta a_B^{*-2}+\frac{1}{2}}\right)\right\}=1,\qquad(6)$$

where $\rho_{a_2}^* = \rho_{a_2}/a_d$.

Our numerical analysis for Eqs. (5) and (6) shows that the magnetic field leads to a considerable change in positions of the terms and to a stabilization of the QW $D_2^-$-states. With the transfer from Eq. (5) to Eq. (6) the role of spatial configuration for the QW $D_2^-$-center becomes apparent: the closeness of QW boundaries for the configuration (5) leads to the energy levels break (cleavage), for degenerated $g$ - and $u$ - states.

## 3. Spectral dependence for the impurity magneto-optical absorption coefficient for the multi-well quantum structure

In this Section, we calculate the impurity magneto-optical absorption coefficient $K_B(\omega)$ for the semiconductive structure, which consists of the tunnelly non-binding QWs taking into account their width dispersion $u = L/\overline{L}$, where $\overline{L}$ is the mean value of the QW width. It is supposed that in every QW there is one $D_2^-$-centre, with two possible spatial configurations, which are described by Eqs. (5) and (6). In the general case, the light absorption coefficient $K_B(\omega)$ can be represented as

$$K_B(\omega) = \frac{2\pi}{\hbar I_0 E_d \overline{L}_c S} \sum_{m=0}^{M} \sum_{n=n_{\min}}^{N} \sum_{n_1=0}^{N_1} P(u^*) \left|M_{f,\lambda}^{(j,k)}\right|^2\bigg|_{u=u^*} \cdot \beta^{*2} \left[\overline{\beta}\left(n+\frac{1}{2}\right)\right]^{-1}, \quad (7)$$

Where $M = [C_1]$ is even part of the number $C_1 = \left(X - \eta_B^2 - (\overline{\beta})^{-1} \cdot u_{\max}^{-1}(n_{\min} + 1/2)\right)/\left(2a_B^{*-2}\right) - 1/2$; $n_{\min} = 0$ or $n_{\min} = 1$ depending on selection rules; $N = [C_2]$ is even part of the number $C_2 = \left(X - \eta_B^2 - a_B^{*-2}(|m| + m + 1)\right)\overline{\beta} u_{\max} - 1/2$; $N_1 = [C_3]$ is even part of the number $C_3 = \left(X - \eta_B^2 - (\overline{\beta})^{-1} \cdot u_{\max}^{-1}(n + 1/2)\right)/\left(2a_B^{*-2}\right) - (|m| + m + 1)/2$; $X = \hbar\omega/E_d$ is the photon energy, in effective Bohr energy units; $I_0$ is the light intensity; $\overline{L}_c$ is the mean value of structure period; $S$ is the area of QW in the plain, which is perpendicular to the growth axis; $u^* = (\overline{\beta})^{-1}(n+1/2)\left(X - \eta_B^2 - a_B^{*-2}(2n_1 + |m| + m + 1)\right)^{-1}$; $\beta^* = \overline{\beta} \cdot u^*$; $u_{\min}, u_{\max}$ are minimal and maximal dispersion values of $u$; $n_1 = 0, 1, 2, \ldots$ is the ra-dial quantum number corresponding to the Landau level; $m = 0, \pm 1, \pm 2, \ldots$ is the magnetic quantum number; $n = 0, 1, 2, \ldots$ is the oscillator quantum number; $\bar{\beta} = \bar{L}^*/(4\sqrt{U_0^*})$; $P(u)$ is the distribution function for the QW width dispersion,

$$P(u) = 2\left[\sqrt{\pi}(\Phi(u_{max} - u_0) + \Phi(u_0 - u_{min}))\right]^{-1} e^{-(u-u_0)^2}, \quad (8)$$

where $\Phi(z)$ is the error function [8]; $u_0 = (u_{min} + u_{max})/2$.

The upper indices $j, k_i$ in matrix element $M_{f,\lambda}^{(j,k_i)}$, which determines the oscillator force value for the dipole optical transition from $g$-state to the state of the QW quasi-discrete spectrum, denote the light polarization direction (with respect to the QW growth axis, $(j = s, t)$) and the molecular ion $D_2^-$ spatial configuration, which is described by Eqs. (5) and (6), $K_1 = (0,0,0 \text{ and } 0,0,z_{a_2})$; $K_2 = (0,0,0 \text{ and } \rho_{a_2}, \varphi_{a_2}, 0)$, correspondingly. For the optical transition with maximal oscillator force $(n_1 = 0, m = 0, n = 0)$, the expression for $M_{f,\lambda}^{(j,k)}$ can be written as

$$M_{f,\lambda}^{(s,k_1)} = -2^{-\frac{5}{4}} \pi^{-\frac{5}{4}} i\lambda_0 \sqrt{\frac{\alpha^* I_0}{\omega}} \gamma B_1 \beta^{\frac{1}{4}} a_B^{-1} a_d^{\frac{1}{2}} \left\{ \left(\beta\eta_B^2 + \beta a_B^{*-2} + \frac{1}{2}\right)^{-1} + \left(\beta\eta_B^2 + \beta a_B^{*-2} + \frac{5}{2}\right)^{-1} \times \right.$$

$$\left. \times 2\pi \exp\left(\frac{z_{a_2}^2}{(2a^2)}\right) \left[ \left(\beta\eta_B^2 + \beta a_B^{*-2} + \frac{1}{2}\right)^{-1} + \left(\beta\eta_B^2 + \beta a_B^{*-2} + \frac{5}{2}\right) \cdot \left(1 - 2\left(\frac{z_{a_2}}{a}\right)^2\right) \right] \right\}, (9)$$

$$M_{f,\lambda}^{(t,k_1)} = -2^{-\frac{1}{4}} \pi^{-\frac{1}{4}} i\lambda_0 \sqrt{\frac{\alpha^* I_0}{\omega}} \gamma B_1 \beta^{\frac{3}{4}} a_B^{-2} a_d^{\frac{3}{2}} \cdot \left(1 + \exp\left(-\frac{z_{a_2}^2}{(2a^2)}\right)\right) \times$$

$$\times \left[ e^{-i\psi} \delta_{m,+1} \left(\beta\eta_B^2 + \beta a_B^{*-2} + \frac{1}{2}\right)^{-1} + e^{i\psi} \delta_{m,-1} \left(\beta\eta_B^2 + 3\beta a_B^{*-2} + \frac{1}{2}\right) \right], \quad (10)$$

$$M_{f,\lambda}^{(s,k_2)} = 2^{-\frac{5}{4}} \pi^{-\frac{5}{4}} i\lambda_0 \sqrt{\frac{\alpha^* I_0}{\omega}} \gamma B_1 \beta^{\frac{1}{4}} a_d^{\frac{1}{2}} \left\{ \left(\beta\eta_B^2 + \beta a_B^{*-2} + \frac{1}{2}\right)^{-1} + \left(\beta\eta_B^2 + \beta a_B^{*-2} + \frac{5}{2}\right)^{-1} + $$

$$+\left(\frac{\rho_{a_2}}{(\sqrt{2}a_B)}\right)^{|m|} \exp\left(-\frac{\rho_{a_2}^2}{(4a_B^2)}\right)\left[\left(\beta\eta_B^2 + \beta a_B^{*-2}(|m|+m+1)+\frac{1}{2}\right)^{-1} + \right.$$

$$\left. +\left(\beta\eta_B^2 + \beta a_B^{*-2}(|m|+m+1)+\frac{5}{2}\right)^{-1}\right]\right\}, \qquad (11)$$

$$M_{f,\lambda}^{(t,k_2)} = -2^{-\frac{1}{4}}\pi^{-\frac{1}{4}}i\lambda_0\sqrt{\frac{\alpha^* I_0}{\omega}}\gamma B_1\beta^{\frac{3}{4}}a_B^{-2}a_d^{\frac{3}{2}} \cdot \left\{e^{-i\psi}\delta_{m,+1}\left(\beta\eta_B^2 + \beta a_B^{*-2} + \frac{1}{2}\right)^{-1} + \right.$$

$$+ e^{i\psi}\delta_{m,-1}\left(\beta\eta_B^2 + 3\beta a_B^{*-2} + \frac{1}{2}\right)^{-1} + \frac{a_B^{*2}}{(2\beta)}\exp\left(-\frac{\rho_{a_2}^2}{(4a_B^2)}\right) \cdot e^{-im\varphi_{a_2}}\left[\Theta(m)(m!)^{-\frac{1}{2}} \times \right.$$

$$\times \left(m\left(\frac{\rho_{a_2}^2}{2a_B^2}\right)^{\frac{m-1}{2}} e^{i(\varphi_{a_2}-\psi)}\left(\nu_0 + m - \frac{1}{2}\right)^{-1} - e^{-i(\varphi_{a_2}-\psi)}\left(\frac{\rho_{a_2}^2}{2a_B^2}\right)^{\frac{m+1}{2}}\left(\nu_0 + m + \frac{3}{2}\right)^{-1}\right) -$$

$$\left. -\Theta(-m-1)(|m|!)^{-\frac{1}{2}}e^{-i(\varphi_{a_2}-\psi)}\left(\frac{\rho_{a_2}^2}{2a_B^2}\right)^{-\frac{m+1}{2}}\left(m + \frac{\rho_{a_2}^2}{2a_B^2}\left(\nu_0 + \frac{3}{2}\right)^{-1}\right)\right]\right\}, \qquad (12)$$

where $\nu_0 = \left(\beta\eta_B^2 + \frac{1}{2}\right)/\left(2\beta a_B^{*-2}\right)$; $\lambda_0$ is the local field coefficient; $\alpha^* = |e|^2/(4\pi\varepsilon_0\sqrt{\varepsilon}\hbar c)$ is the fine structure constant which accounts for the static relative dielectric permeability $\varepsilon$; $c$ is the speed of light in vacuum; $\delta_{i,k}$ is Kronecker symbol; $\Theta(x)$ is Havyside function;

$$B_1 = \left(\pi^{\frac{3}{2}}a_B^2 a E_d^2/(\gamma^2\beta^2)\right)^{\frac{1}{2}}\left[\sqrt{\pi}\sum_{k=0}^{\infty}\Gamma\left(\frac{d}{2}\right)G\Psi(d)\middle/\left(4\Gamma\left(\frac{d+1}{2}\right) + \beta^{-2}a_B^{*4}\exp\left(-\frac{\rho_{a_2}^2}{(4a_B^2)}\right)\right)\right] \times$$

$$\times \Gamma\left(\frac{\beta\eta_B^2 + 1/2}{\beta a_B^{*-2}}\right)G_2\left(\left(\frac{\beta\eta_B^2 + 1/2}{\beta a_B^{*-2}}\right),0,\left(\frac{\rho_{a_2}^2}{4a_B^2}\right)\right)\right]^{-1/2}$$; $G_2(\alpha,\gamma,z)$ is the degenerate hypergeometric function of the second kind [8]; $\Psi(x)$ is the logarithmic derivative of Euler gamma-function; $d = \beta\eta_B^2$; $G \approx 0,916$ is the Katalane constant [8]; $\psi$ is the polar angle for the polarization unit vector $\vec{e}_\lambda$ in the cylindrical system of reference.

Figures 1 (a, b) and 2 (a, b) show the spectral dependences $K_B^{(t,k_1)}(\omega)$, $K_B^{(s,k_1)}(\omega)$ and $K_B^{(t,k_2)}(\omega)$, $K_B^{(s,k_2)}(\omega)$, which are correspondingly calculated due to Eq. (7) for the multi-well quantum structure based on InSb. As seen from Figs. 1 and 2, the change of direction of the light polarization leads to a drastic modification of the profile of absorption spectral curve (compare (a) and (b) in Figs. 1 and 2). This is partially related with the change of selection rules for the magnetic quantum number. From a comparison of Figs. 1(a) and 2(a), and Figs. 1(b) and 2(b) one can observe an essential role of the spatial configuration for the QW $D_2^-$ molecular ion. Namely, one can see not only spatial curve profile, but also the absorption value too.

In summary, we have shown that the magneto-optical absorption anisotropy in multi-well quantum structure is related with not only the light polarization direction but also with the $D_2^-$-ion spatial configuration.

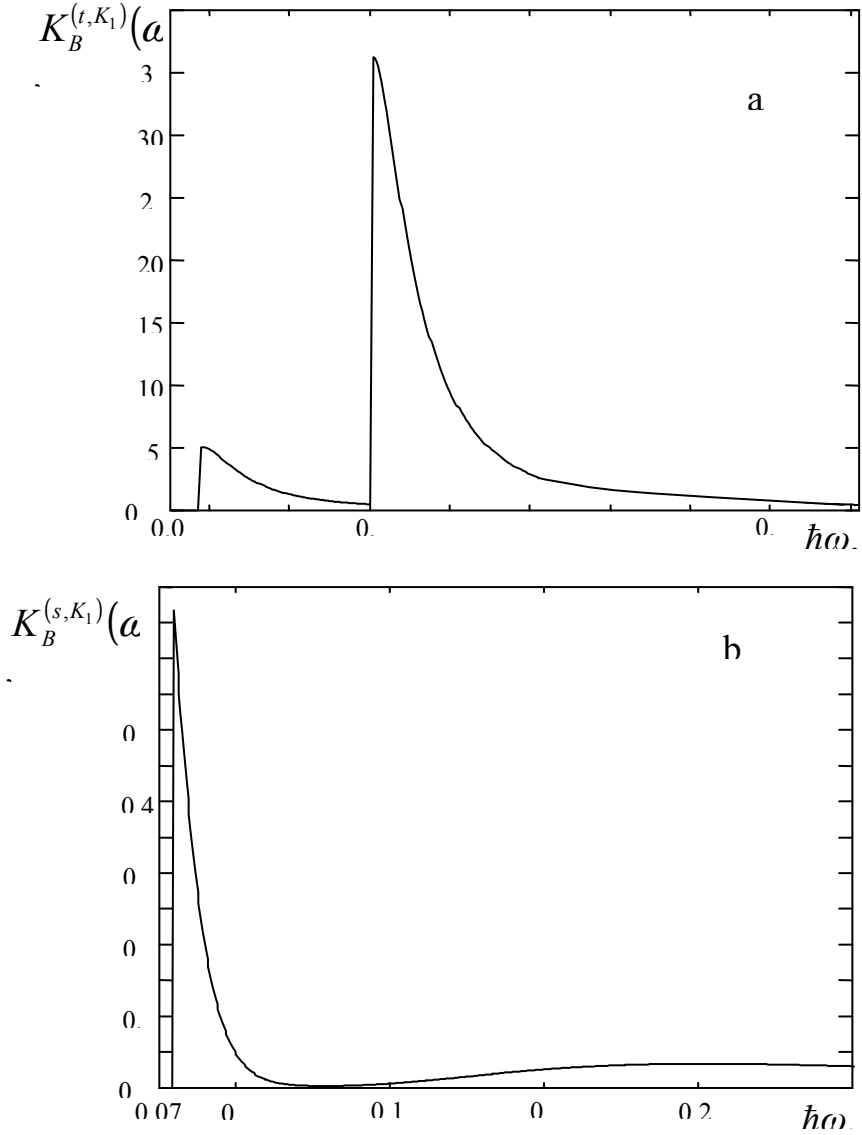

**Fig. 1.** The spectral dependence for the magneto-optical impurity absorption coefficient: (a) the transversal polarization case $\left(K_B^{(t,K_1)}(\omega)\right)$; (b) the longitudinal polarization case $\left(K_B^{(s,K_1)}(\omega)\right)$; $|E_i| = 5{,}5 \cdot 10^{-2}\, eV$, $\bar{L} = 71{,}6\, nm$, $U_0 = 0{,}2\, eV$, $z_{a_2}^* = 0{,}25$, $B = 5\, T$, $\vec{R}_{a_1} = (0,0,0)$, $\vec{R}_{a_2} = (0,0,z_{a_2})$.

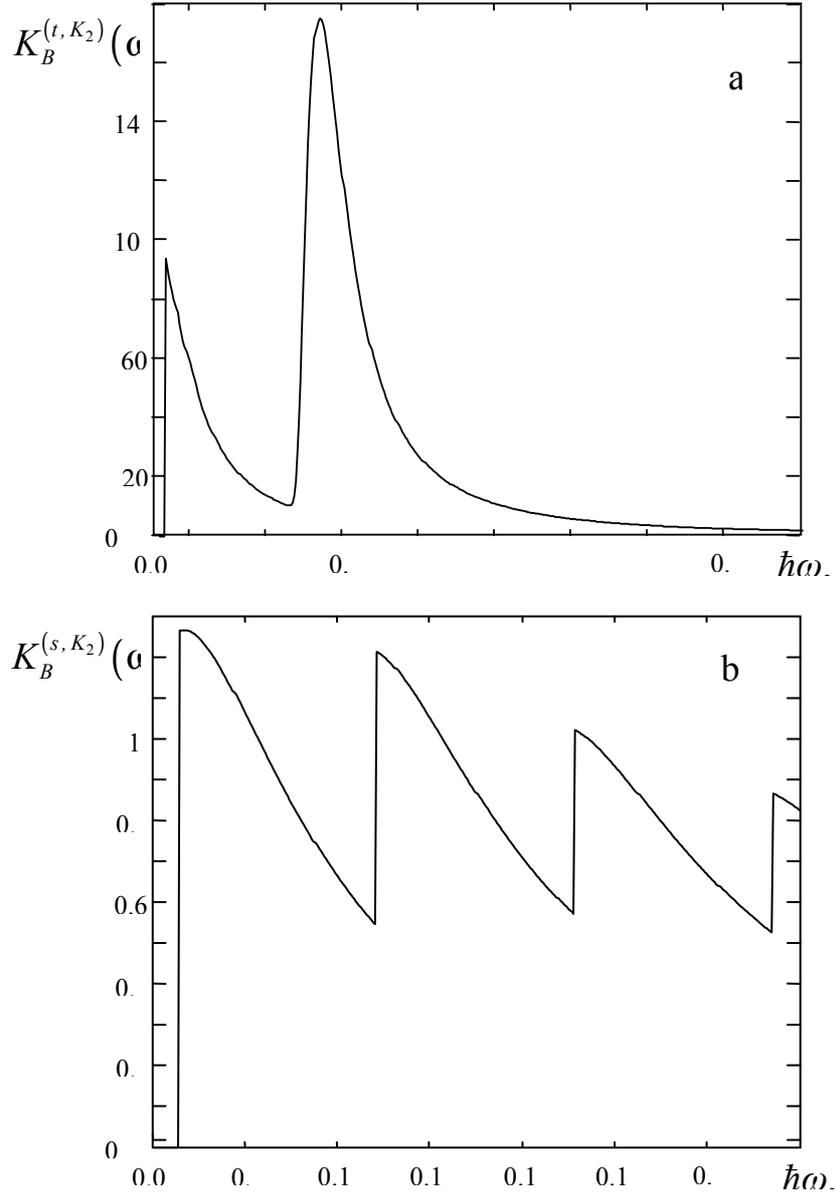

**Fig. 2.** The spectral dependence for the magneto-optical impurity absorption coefficient: (a) the transversal polarization case $(K_B^{(t,K_2)}(\omega))$; (b) the longitudinal polarization case $(K_B^{(s,K_2)}(\omega))$; $|E_i| = 5.5 \cdot 10^{-2}\,eV$, $\bar{L} = 71.6\,nm$, $U_0 = 0.2\,eV$, $\rho_{a_2}^* = 0.25$, $B = 5\,T$, $\vec{R}_{a_1} = (0,0,0)$, $\vec{R}_{a_2} = (\rho_{a_2}, \varphi_{a_2}, 0)$.